\documentclass[%
 reprint,
 amsmath,amssymb,
 aps,
]{revtex4-2}

\usepackage{graphicx}% Include figure files
\usepackage{dcolumn}% Align table columns on decimal point
\usepackage{bm}% bold math
\usepackage{amsmath}% bold math
\usepackage{mathtools}% bold math

\usepackage[dvipsnames]{xcolor}
\begin{document}
%%% DOUBLE EQUATION %%%
\makeatletter
\newcommand*\@dblLabelI {}
\newcommand*\@dblLabelII {}
\newcommand*\@dblequationAux {}

\def\@dblequationAux #1,#2,%
    {\def\@dblLabelI{\label{#1}}\def\@dblLabelII{\label{#2}}}

\newcommand*{\doubleequation}[3][]{%
    \par\vskip\abovedisplayskip\noindent
    \if\relax\detokenize{#1}\relax
       \let\@dblLabelI\@empty
       \let\@dblLabelII\@empty
    \else % we assume here that the optional argument
          % has the required shape A,B
       \@dblequationAux #1,%
    \fi
    \makebox[0.5\linewidth-1.5em]{%
     \hspace{\stretch2}%
     \makebox[0pt]{$\displaystyle #2$}%
     \hspace{\stretch1}%
    }%
    \makebox[0.5\linewidth-1.5em]{%
     \hspace{\stretch1}%
     \makebox[0pt]{$\displaystyle #3$}%
     \hspace{\stretch2}%
    }%
    \makebox[3em][r]{(%
  \refstepcounter{equation}\theequation\@dblLabelI, 
  \refstepcounter{equation}\theequation\@dblLabelII)}%
  \par\vskip\belowdisplayskip
}
\makeatother
%%% END DOUBLE EQUATION %%%

\preprint{APS/123-QED}

\title{Using Activity to Compartmentalize Binary Mixtures}% Force line breaks with \\
%\thanks{A footnote to the article title}%

\author{Nicholas J Lauersdorf}
\affiliation{%
 Department of Applied Physical Sciences, University of North Carolina, Chapel Hill, North Carolina 27515, USA.}%
 
\author{Ehssan Nazockdast}%
 \email{ehssan@email.unc.edu}
\affiliation{%
 Department of Applied Physical Sciences, University of North Carolina, Chapel Hill, North Carolina 27515, USA.} 

\author{Daphne Klotsa}%
 \email{dklotsa@unc.edu}
\affiliation{%
 Department of Applied Physical Sciences, University of North Carolina, Chapel Hill, North Carolina 27515, USA.} 
 
%\linebreak
%\date{\today}% It is always \today, today,
             %  but any date may be explicitly specified
\begin{abstract}
We computationally study suspensions of slow and fast active Brownian particles that have undergone motility induced phase separation and are at steady state. Such mixtures, of varying non-zero activity, remain largely unexplored even though they are relevant for a plethora of systems and applications ranging from cellular biophysics to drone swarms. Our mixtures are modulated by their activity ratios  ($\mathrm{Pe}^\mathrm{R}$), which we find to encode information by giving rise to three regimes, each of which display their unique emergent behaviors. Specifically, we found non-monotonic behavior of macroscopic properties, \textit{e.g.} density and pressure, as a function of activity ratio, microphase separation of fast and slow particle domains, increased fluctuations on the interface and severe avalanche events compared to monodisperse active systems. Our approach of simultaneously varying the two activities of the particle species allowed us to discover these behaviors and explain the microscopic physical mechanisms that drive them.

\begin{description}
\item[DOI]
Secondary publications and information retrieval purposes.

\end{description}
\end{abstract}

%\keywords{Suggested keywords}%Use showkeys class option if keyword
                              %display desired
\maketitle

%another way to condense first and second paragraphs:
\noindent \textit{Introduction.} Active matter describes nonequilibrium systems comprised of components that locally consume energy to move, such as swarms of bacteria and clusters of self-propelled colloids. Past studies on monodisperse systems and minimal models have led to the discovery of a range of remarkable emergent phenomena including motility-induced phase separation (MIPS)~\cite{Theurkauff2012, Redner2013a, Cates2015a, Ridgway2023, Zhao2023, Caprini2020, VanDerLinden2019, Digregorio2018, Ro2022, Caporusso2020}, swarming and flocking \cite{Vicsek1995, Romanczuk2008, Sarkar2021, Zhang2022, Chen2024a, Caprini2023, Yu2022, Maity2023, Benvegnen2023}, and pattern formation~\cite{Liebchen2017, Tiribocchi2015, Dey2012, Kursten2020, Linkmann2019, Dombrowski2004, Dunkel2013, Gachelin2014, Duan2023, Chakrabarti2023, Colombo2023, Reinken2022, Peng2016a, Albers2023, Faluweki2023}. 
Adding complexity, studies of polydisperse mixtures have included components of different sizes~\cite{Yang2014a, McCandlish2012}, shapes~\cite{Nguyen2014, Moran2022}, interactions~\cite{Bartnick2016, Chepizhko2013, Chen2024a, AgudoCanalejo2019, Duan2023, Dinelli2023, Alston2023, Ivlev2015, Saha2020}, and propulsion mechanisms~\cite{Levis2019, Yeo2015, Nguyen2014, Bardfalvy2020, Chen2024a, Ai2018, Kole2021, Caprini2023, Arora2021}. The relative activities of the particles have mostly been varied in the limiting case of active/passive mixtures~\cite{Stenhammar2015, Takatori2015d, Wittkowski2017a, Dolai2018a, VanDerMeer2020a, Wysocki2016a, Agrawal2021, RogelRodriguez2020a, Bhattacharyya2023, Hecht2022, Caballero2022}, passive particles in active baths~\cite{Morozov2014, Pushkin2013, Lin2011, Ishikawa2010, Underhill2008, Preisler2016, Granek2022, Peng2016a, Wang2020, Angelani2011, Maitra2020, Angelani2009, Liu2020, Malgaretti2022, Saha2023, Maes2020}, and active particles with static obstacles~\cite{Huang2017a, Nayak2023, Chepizhko2013, Duzgun2018, Das2020, Forgacs2021, Nikola2016, Codina2022, Schimming2024, Leite2016, Borba2020, VanRoon2022, Martinez2020, Spagnolie2015, Martinez2018}. 
%(Note that the term passive refers to particles undergoing Brownian motion). 
However, mixtures of particles with distinct nonzero active driving forces remain largely unexplored, even though they are relevant in a variety of systems ranging from fundamental physics to applications, as we explain next.
Recent studies are an exception and show there is growing interest~\cite{Kolb2020, Decastro2021a, Decastro2021b, RojasVega2023, Debnath2020, Rojas-Vega2023a}.

Studying active/active mixtures is interesting from a fundamental physics standpoint, since it is a simple nonequilibrium model system that challenges our understanding of what collective behavior to expect, and what quantities equilibrate; still it is sufficiently simple to allow theoretical progress from first principles. We often think of activity as a temperature analogue. But while fast particles can effectively exchange activity with slower ones when in contact, as has been observed in active/passive~\cite{Wilson2011, Chen2007, Wu2000, Mino2013, Jepson2013, Valeriani2011, Leptos2009, Kurtuldu2011, Kasyap2014, Restrepo-Perez2014a, Angelani2011, Granek2022, Mino2011, Baek2018, Forgacs2021, Wang2020} and slow/fast~\cite{Debnath2020} systems, they do not reach a uniform activity and so the answer is unclear. In biology there are many active matter systems of importance that are relevant to biophysics and medicine, and they are typically polydisperse including in their motility speeds~\cite{Gough2018, Knight2003, Heck1981, Jolles2017, Jolles2020, Klamser2021, Christensen1996, Forget2022, Jerison2020}, such as bacteria and sperm cells. 
Furthermore in synthetic systems, limitations in fabrication techniques give rise to a distribution of swim speeds in self-propelled colloids~\cite{Palacci2010, Palacci2013, Ebbens2010, Lee2014}, but there is limited understanding of how that affects the emergent behavior. Finally from a soft-matter and materials physics perspective, tuning the relative activities of particles opens an enormous parameter space 
%for controlling the design of complex assemblies
with potentially a wealth of physics providing opportunities for more control in designing dynamic complex assemblies.
Active matter mixtures exhibit behaviors akin to those found in living systems \textit{e.g.} nonequlibrium transitions, microphase separation, and bistable states, that have not been observed before. 

%that can lead to models forinformation transfer and physical learning, etc. 

In this paper, we performed simulations of binary active mixtures of fast and slow Active Brownian Particles (ABPs), that have undergone motility induced phase separation (MIPS) and are in steady state. We studied the properties of the phase separated system as a function of the ratio of the slow to fast particle activities, $0\le \text{Pe}^\mathrm{R} \le 1$, and discovered three regimes corresponding to small, intermediate and large activity ratios. 
At large activity ratios the fast and slow particles are uniformly mixed and the behavior is analogous to monodisperse ABP suspensions, even when the slow activity is nearly a third of the fast. 
At intermediate and small activity ratios, when the particles are increasingly heterogeneous in their activities, is where we saw the most interesting behavior: 
\textit{Microscopically}, the system exhibits microphase separation, increased avalanche events and fluctuations, and active herding (slow particles pushed by the fast ones). We also found nonmonotonic behavior in \textit{macroscopic} quantities, including cluster pressure, density and compressibility.  
To obtain a deeper understanding, we developed a coarse-grained continuum model, which provided further insight and whose predictions were in good agreement with the simulation results. We thus propose a physical mechanism that links microscopic and macroscopic behavior and explains the observed emergent phenomena.

\begin{figure}[h!]
\includegraphics[width=0.48\textwidth,trim={0 0 0 0},clip]{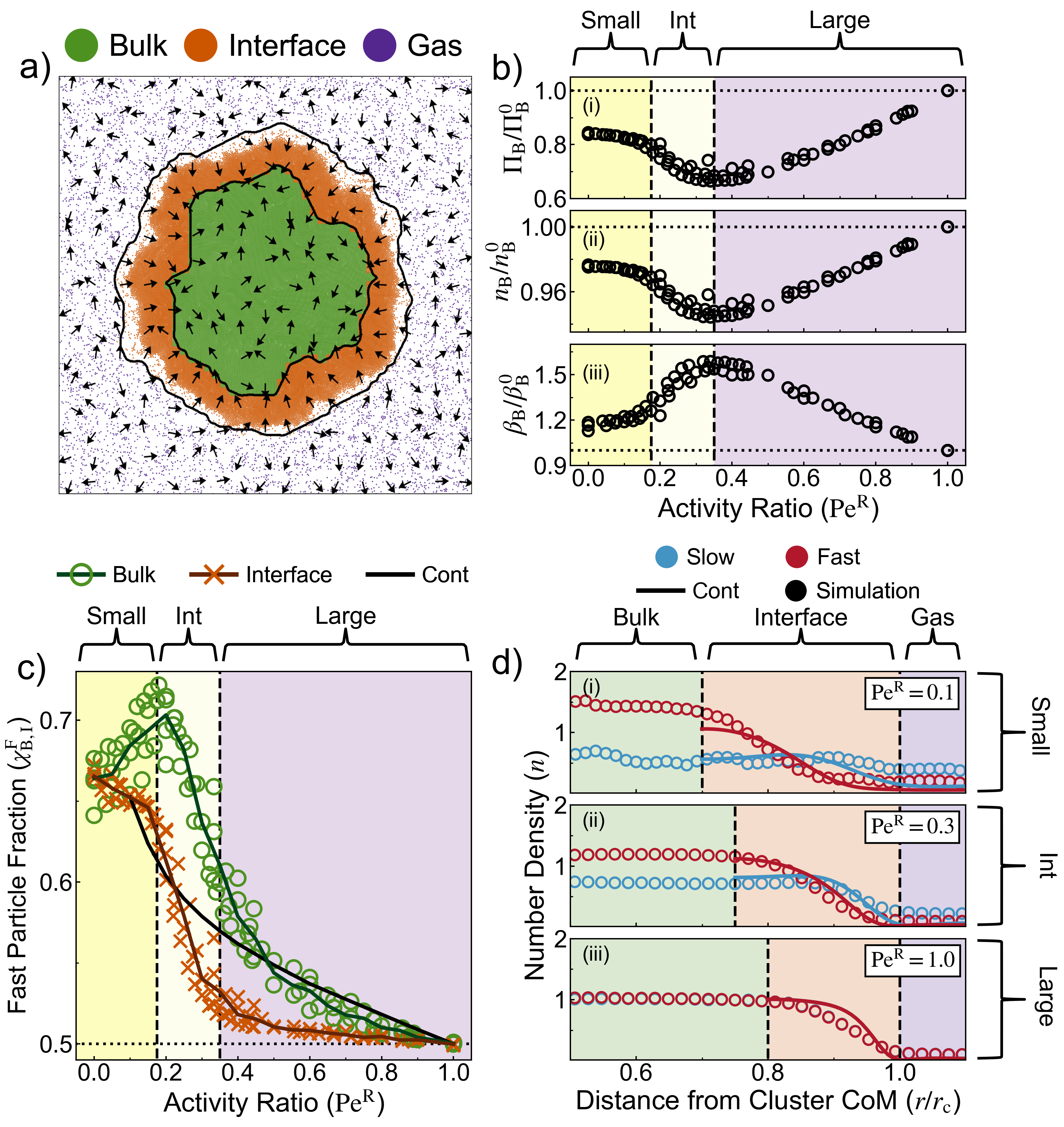}% Here is how to import EPS art
\caption{\label{fig:one} a) A simulation snapshot of a monodisperse ABP system at steady state after MIPS with $\mathrm{Pe} = 500$. Phases are labeled bulk (green), interface (orange), and gas (purple) and average particle orientation is shown as arrows. b) Bulk (cluster) (i) interparticle pressure, (ii) number density, and (iii) compressibility normalized by their respective values in monodisperse (same activity, $\mathrm{Pe}^\mathrm{R}=1$) case, as a function of activity ratio ($\mathrm{Pe}^\mathrm{R}$). c) The steady-state fast particle fraction in the bulk phase ($\chi^\mathrm{F}_\mathrm{B}$, green circle) and the interface ($\chi^\mathrm{F}_\mathrm{I}$, orange cross) as a function of activity ratio ($\mathrm{Pe}^\mathrm{R}$). The solid black line represents the 1D-continuum model predictions. In (b) and (c), three regimes, corresponding to three range of activity ratios, are visualized by background color and separated by dashed black lines. d) Number density of slow (blue) and fast (red) particles for our ABP simulation (dots) and 1D continuum model (solid line) as a function of distance from the cluster's center of mass ($r$) normalized by the cluster radius ($r_\mathrm{c}$) for $\mathrm{Pe}^\mathrm{R}=0.1$, $0.3$, and $1.0$. 
}
\end{figure}

 \noindent \textit{Model \& Methods:}
Using HOOMD-Blue~\cite{Anderson2020, Glaser2015, Anderson2008}, we simulated ABPs of diameter $\sigma$ that translate and rotate over time (measured in units of the Brownian diffusion time, $\tau_\mathrm{B}$) in accordance with overdamped Langevin dynamics and interact via the Weeks-Chandler-Anderson (WCA) potential with a soft repulsive well depth $(\epsilon=1)$. Each particle is subject to an active force, ${\mathbf{F}}_\text{a}=\xi v_\text{p}\widehat{\mathbf{p}}$, where $v_\mathrm{p}$ is the swim speed in the absence of collisions and it modulates the activity, quantified through the Peclet number ($\mathrm{Pe}=3v_\text{p}\tau_\text{r}/ \sigma$ where $\tau_\mathrm{r}$ is the characteristic reorientation time). We simulate mixtures of fast ($\mathrm{Pe}^\mathrm{F}$) and slow ($\mathrm{Pe}^\mathrm{S}$) ABPs whose activities are varied independently between 0-500 with particle fractions
$\chi^\mathrm{F}$ and $\chi^\mathrm{S}$ respectively, where 
$\chi^\mathrm{F}=1.0-\chi^\mathrm{S}$, varied between 0.2-0.8 at $\phi=0.6$. We ran a range of relative fractions (see SI \S 1); in the main paper for simplicity we show results for $\chi^\mathrm{F}=\chi^\mathrm{S}=0.5$, but the findings remain the same. We define the activity 
ratio $\mathrm{Pe}^\mathrm{R}=\frac{\mathrm{Pe}^\mathrm{S}}{\mathrm{Pe}^\mathrm{F}}$, ranging from a monodisperse active system ($\mathrm{Pe}^\mathrm{R}=1$) to the most heterogeneous ($\mathrm{Pe}^\mathrm{R}=0$) which is a mixture of passive and active particles (Fig. S1). It is useful to also define the net activity, $\mathrm{Pe}^\mathrm{Net}=\chi^\mathrm{S}\mathrm{Pe}^\mathrm{S}+\chi^\mathrm{F}\mathrm{Pe}^\mathrm{F}$~\cite{Kolb2020}, varied between 0-500. More details about the model and the simulation parameters can be found in the SI \S 1.

We studied systems that have undergone MIPS and have reached steady state.
We ran simulations using a variety of initial conditions, \textit{e.g.} random gas and differently-instantiated clusters, and for very long times ($600\tau_B$); unlike previous works~\cite{Stenhammar2015} we found that all initializations eventually led to the same steady-states for $\mathrm{Pe}^\mathrm{R}\ge0.2$. For $\mathrm{Pe}^\mathrm{R}<0.2$, the large fluctuations in the cluster size prolong MIPS and make it much less likely; hence, for low activity ratios we instantiated the clusters (and then observed them melt or persist over long times). We provide a detailed discussion in SI \S 2. We distinguish three phases shown in Fig.~\ref{fig:one}(a): 1) a dilute \emph{gas} phase with random alignment of active forces; 2) a \emph{bulk} with uniform high density and also random alignment of active forces; and 3) the \emph{interface}, that sits between the bulk and the gas, and is characterized by a transition from high to low density and a net alignment of particles' active forces and a net compressive force towards the center of the cluster~\cite{Lauersdorf2021a} (SI \S 3). 

%\section{Nonmonotonic mechanical properties}
\textit{Results:} We begin by studying the macroscopic mechanical properties of the bulk phase, specifically the average pressure as a function of the activity ratio, $\text{Pe}^\text{R}$. In monodisperse systems, pressure has a nearly  linear dependence with activity~\cite{Omar2019, Lauersdorf2021a}. To remove this dependence and focus on the activity ratio, for each system we normalize the pressure by the computed pressure at activity ratio $1$, denoted $\Pi^0_\mathrm{B}$. This non-dimensionalization collapses all the data from different net activities into a single curve (Figs.~\ref{fig:one}(b)(i), S2). As shown, the dimensionless pressure displays \emph{nonmonotonic} variations with activity ratio with a minimum around $\text{Pe}^\text{R} \approx 0.35$, followed by an increase and then a plateau at small activity ratios, $\text{Pe}^\text{R}<0.175$. The number density of particles in the bulk phase ($n_\mathrm{B}$) and compressibility ($\beta_\mathrm{B}=\frac{1}{n_\mathrm{B}}\frac{\partial n_\mathrm{B}}{\partial \Pi_\mathrm{B}}$) also show non-monotonic behavior with $\text{Pe}^\text{R}$ similar to the pressure, (Fig.\ref{fig:one}(b)(ii) and (iii), and S2).
From these observations, we can define three regimes in terms of activity ratio: large $0.35<\text{Pe}^\text{R}\le 1$, intermediate $0.175<\text{Pe}^\text{R}<0.35$ and small $\text{Pe}^\text{R}<0.175$. These regimes emerge naturally and consistently from our results obtained from independently measured quantities besides pressure, density and compressibility, as will be shown in this paper.

\begin{figure*}[t!]
\includegraphics[width=0.84\textwidth,trim={0 0 0 0},clip]{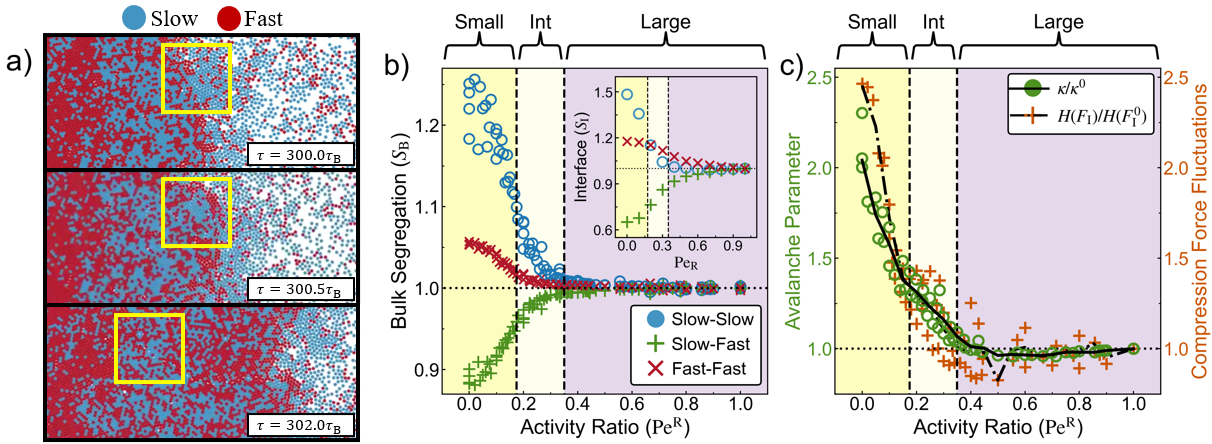}% Here is how to import EPS art
\caption{\label{fig:three} a) Simulation snapshots of an active/passive mixture zoomed-in on the interface showing the formation of a segregated domain of passive particles from the interface (top row) and its incorporation into the bulk (two bottom rows). b) The degree of segregation for slow-slow, slow-fast, and fast-fast neighbor pairs within the bulk as a function of activity ratio ($\mathrm{Pe}^\mathrm{R}$). The inset shows segregation at the interface; see also SI section \S 5. c) Avalanche parameter ($\kappa$, green circle) and compression force fluctuations ($H(F_\mathrm{I})$ (defined in SI \S 6), orange cross) normalized by their respective values in monodisperse (same activity, $\mathrm{Pe}^\mathrm{R}=1$) case, vs ratio ($\mathrm{Pe}^\mathrm{R}$).}
\end{figure*}

To explore the microscopic basis of this nonmonotonic behavior, we computed the relative composition of fast and slow particles in the cluster (Figs.~\ref{fig:one}(c) and S3). Note that similar to the dimensionless pressure, the fraction of fast (and slow) particles is only a function of activity ratio and independent of the net activity. 
We observe that as $\text{Pe}^\text{R}$ decreases the fast particle fraction increases in the bulk and the interface, particularly for intermediate and low $\text{Pe}^\text{R}$ (Figs.~\ref{fig:one}(c), S3). 
At the interface, because the active forces are aligned, an increase of the number of fast particles leads to a larger compression force pressing towards the bulk. Hence, the otherwise counterintuitive uptick in the bulk pressure at intermediate $\mathrm{Pe}^\mathrm{R}$ (Fig.~1(b)(i)) may result from this increase in compression force. However, it remains unclear why there are more fast particles in the cluster in the intermediate and low $\mathrm{Pe}^\mathrm{R}$ regimes. To search for answers, next we considered the spatial distribution of fast and slow particles.

As shown in Fig. \ref{fig:one}(d), for large activity ratios, including the monodisperse case $\text{Pe}^\text{R}=1$, the number density of fast and slow particles is nearly equal and increases across the interface from a low uniform value in the gas to a large uniform value in the bulk. For intermediate and low activity ratios, however, we see a split in how fast and slow particles are distributed. At the interface, the particle fraction radially transitions from majority slow particles nearest to the gas to
majority fast nearest to the bulk, similar to active/passive MIPS~\cite{Wysocki2016a} and slow and fast ABPs assembling on a rigid wall~\cite{Rojas-Vega2023a}. This observation can be understood by thinking of the dense cluster as a rigid boundary, on which fast and slow particles assemble. In the simplest case of a dilute suspension of fast and slow point particles, the density variations with distance from the rigid boundary (bulk surface), $x$, simplify to: $n^\mathrm{S,F} \approx n^\mathrm{S,F}_0+E \exp{(-2 x\mathrm{Pe}^\text{(S,F)}/\sigma)}$, where $n_0$ is the net (average) density of slow or fast particles in the gas phase, and $E$ is an integration constant determined by boundary conditions; see SI \S 4 for details. 
As a result, we expect the fast particles to concentrate on the interior of the cluster and form a thinner boundary layer, compared to slow particles, which is in agreement with simulation results in Fig~\ref{fig:one}(d). 

A close inspection of simulation movies revealed that instead of two distinct radial layers, we observe microphase separated domains of fast and slow particles at the interface, which get integrated into the bulk as well (Fig~\ref{fig:three}(a)).
To quantify microphase separation as a function of activity ratio, we computed the pair probability density of
observing slow-slow, slow-fast and fast-fast pairs as nearest neighbors in both the bulk and interface normalized by the average slow or fast particle fraction of the respective phase (Fig~\ref{fig:three}(b)).
At large activity ratios, the particles are spatially uniform and bulk or interface segregation goes to 1.
At intermediate and low activity ratios, 
we observe a significant increase in the probability of observing slow-slow and to a lesser degree fast-fast pairs and, therefore, lower probability of observing slow-fast pairs.
We are thus showing that microphase separation, reported previously mainly for active/passive~\cite{Stenhammar2015, Dolai2018a, Kolb2020, Yang2014a, Agrawal2021, Wysocki2016a, McCandlish2012}, is a function of activity ratio and begins much earlier in the intermediate $\mathrm{Pe}^\mathrm{R}$ regime.

Simulation results indicate that microphase separated domains start at the interface and move into the bulk (Fig.~\ref{fig:three}(a)).  
To test this, we developed a 1D coarse-grained model that solves for the time-dependent variations in the number density and average alignment of two different active species in ABP suspensions. The only ingredients of the model are thermal diffusion, activity, and interparticle forces; details are provided in the SI \S4. 
Interestingly, the model predicted that during the early stages of MIPS,
at intermediate activity ratios, the interface and the bulk phase are primarily composed of fast particles. With time, the density of fast particles is reduced, while the density of slow particles is increased. This supports that microphase separation starts at the interface and gets incorporated into the bulk, see also (SI \S 4). 
Furthermore, the results of the model are in agreement with the computed splitting of slow and fast particles' number density at the interface for low and intermediate activity ratios shown in Fig~\ref{fig:one}(a)(i) and (ii). The model also correctly predicts the enrichment of fast particles in the dense phase at low and intermediate activity ratios.

In monodisperse systems the active particles leave the interface when they orient away from the cluster. The desorption timescale is, thus, determined by the rotational diffusion timescale of the particles.  
However, in simulations it has been shown that the desorption of a particle typically coincides with large groups of particles collectively escaping from the cluster via \textit{avalanche-like events} ~\cite{Redner2013a, Redner2013, Soto2014, Reichhardt2018, Reichhardt2018a, Lauersdorf2021a, Rojas-Vega2023a}. We expect that in our active mixtures the presence of microphase-separated domains especially at the interface would lead to stronger fluctuations in the compression force (that holds the cluster together) and thus avalanche events.  
%These avalanche events result from the large disturbances in the local pressure induced by a few particles leaving the cluster. 
%We reasoned that the structural heterogeneity that results from microphase separation especially at the interface also leads to stronger spatial and temporal fluctuations in the active compression force at the interface, leading to stronger avalanche events.

\begin{figure}[t!]
\includegraphics[width=0.3\textwidth,trim={0 0 0 0},clip]{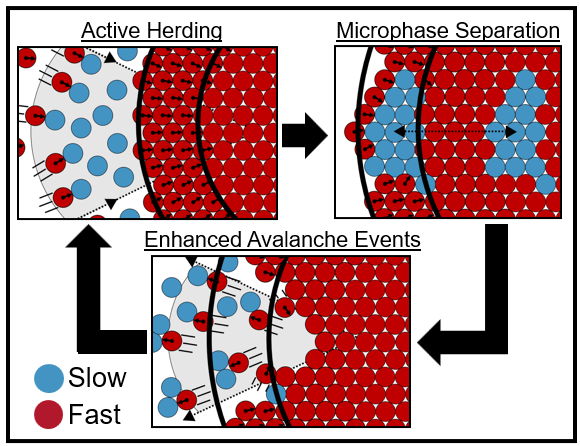}% Here is how to import EPS art
\caption{\label{fig:four} Schematic of the proposed mechanism showing the process of active herding, microphase separation, and enhanced avalanche events.}
\end{figure}

To test this hypothesis, we turned to our ABP simulations and computed the desorption rate of 
particles from the cluster at different activity ratios. The avalanche parameter is defined as the ratio of the desorption rate from simulations to that predicted from rotational diffusion time, i.e. larger values denote stronger avalanche events. 
We also computed the fluctuations around the average compression force at the interface, integrated over the thickness of the interface and over time, at different activity ratios (Fig~\ref{fig:three}(c)). Both quantities are normalized by their corresponding values in the monodisperse system to study the effect of activity ratio. 
We find that both the avalanche parameter and the compression force fluctuations trace each other over the entire range of activity ratios. 
%indicating that they are correlated. 
%and drive each other: fluctuations in the compression force lead to avalanches and vice versa. 
%demonstrating that avalanche events are driven by fluctuations in compressive force (or local pressure).
More importantly, both parameters closely follow the trend observed in the bulk (and interface) segregation parameter: they remain nearly unchanged at larger activity ratios and show a sharp increase at intermediate and low activity ratios. 
These observations support the following mechanism: microphase separation at the interface leads to 
increases in the fluctuations of compression  and much stronger avalanche events, where up to half of the particles in the cluster escape (Fig. S4). In contrast, at high activity ratios fewer than $2\%$ of the cluster particles escape during avalanche events (Fig. S4).

The steady state microstructure is determined by the balance of adsorption and desorption of each particle species. 
%At steady state the adsorption and desorption rates are balanced, 
Thus, we expect the strong avalanche events at intermediate and low $\mathrm{Pe}^\mathrm{R}$ (i.e. high desorption), to be balanced by enhanced adsorption. Indeed, simulations show that  
fast particles push and  
deposit slow particles from the gas onto the interface, which we call \emph{active herding} (SI \S7). As they do so, the fast particles push through the slow layer at the interface and assemble closer to the bulk boundary or move into the bulk phase.

Based on these observations, we propose the following mechanism for the steady state of binary active mixtures at intermediate and low $\mathrm{Pe}^\mathrm{R}$: MIPS begins with the nucleation of small clusters of fast particles. These clusters act as rigid boundaries on which slow and fast particles adsorb. Fast particles herd slow particles leading to slow enhanced adsorption. The cluster overall has more fast particles than slow, while the net density remains roughly unchanged with activity ratio.  This leads to the nonmonotonic behavior in the macroscopic quantities (pressure etc). The difference in activities leads to microphase separation of slow and fast particles at the interface, with fast particles assembling at the inner layer closer to the bulk and the slow particles assembling at the outer layer of the interface closer to the gas phase. 
As more particles assemble around the existing interface, the cluster grows in size and the phase-separated slow and fast particles that were making up the interface at earlier times  get integrated to the cluster. 
The microphase-separated domains at the interface lead to larger fluctuations in the compression force which in turn leads to enhanced avalanche events. 
The process repeats with the re-adsorption of fast and slow particles from the gas onto the interface through active herding.

We have shown that our approach of continuously varying the activity of particle species revealed the existence of three regimes of P\'eclet ratio, previously unknown. The regimes that emerge are robust and demonstrate several distinct structural and dynamical features in micro- and macro-scale. Thus the behavior of mixtures is not a simple interpolation between the active monodisperse case and active/passive; rather there is a richness in between the extremes. Based on our results, we propose a mechanism that describes the dynamic steady-state of the system at low and intermediate $\mathrm{Pe}^\mathrm{R}$ through a series of steps that include interesting nonequilibrium phenomena including active herding, microphase separation, active force fluctuations at the interface and strong avalanche events. We emphasize the role of the interface, which compartmentalizes the bulk, acting like a semi-permeable membrane that selectively filters out slow particles and allows fast particles in. In other words, the interior structure of the cluster is modulated by the structure and mechanics of the interface. 
%(SI \S8, \S9)
%Any active system, can be treated as surface to template the interior structure.
%any active system, can treat as surface can be used to template the interior structure
%ie if we can control/affect the interface -> we can affect the interior? can we say that.?

\textit{Acknowledgments.} This work was generously made possible by the Department of Defense (DoD) National Defense Science and Engineering Graduate (NDSEG) Research Fellowship.

\bibliography{binary_active_paper}% Produces the bibliography via BibTeX.

\end{document}